\documentstyle[preprint,aps]{revtex}
\begin{document}
\draft
\tighten

\title{Resistance scaling at the Kosterlitz--Thouless transition}

\author{Mats Wallin{$^1$} and Hans Weber{$^2$}}

\address{{$^1$}Department of Theoretical Physics,
Royal Institute of Technology, S--100 44 Stockholm, Sweden}

\address{{$^2$}Department of Physics, Lule\aa \ University of Technology,
S--971 87 Lule\aa , Sweden.}

\date{\today}
\maketitle
\widetext

\begin{abstract}
We study the linear resistance at the Kosterlitz--Thouless transition
by Monte Carlo simulation of vortex dynamics.
Finite size scaling analysis of our data
show excellent agreement with scaling properties of the
Kosterlitz--Thouless transition.
We also compare our results for the linear resistance with experiments.
By adjusting the vortex chemical potential to an optimum value,
the resistance at temperatures above the transition temperature
agrees well with experiments over many decades.
\end{abstract}
\pacs{PACS numbers: 05.70Jk, 64.60Cn, 74.20De}
\narrowtext

Scaling of current--voltage characteristics is useful both in
theoretical and experimental analysis of superconductors.
One useful way to calculate linear and nonlinear resistance
accurately is by computer simulation, and scaling analysis is crucial to
extract the critical properties.
The linear resistance and nonlinear current--voltage
characteristic are key quantities in the
recently much studied vortex glass state in
disordered superconductors~\cite{VG}.
At the vortex glass transition the nonlinear current--voltage
characteristic is universal, and the linear resistance vanishes,
signaling true superconductivity.
The current--voltage characteristic is also a key quantity
of the zero--field superconducting transition in two dimensions,
which is a Kosterlitz--Thouless (KT) transition~\cite{KT,Bere}.
The resistance curve above the KT transition is given by an (unknown)
universal function of a reduced Coulomb gas temperature variable,
allowing experimental data for different 2D systems to
collapse on the same curve, as shown by Minnhagen~\cite{PM}.
Remarkably, also certain 3D high temperature superconductors,
with weakly coupled layers,
seem to fit the same curve, except very close to $T_c$~\cite{PM1,HTSexp}.
The term ``universality'' for the resistance curve is used in the
same sense as e.g.\ the universality of the BCS gap function,
rather than in the sense of critical phenomena.
The form of the resistance curve has been accurately constructed from
such experiments, but an accurate calculation is lacking.
KT theory gives an approximate expression
for the resistance, valid in the KT critical region,
but which cannot accurately fit the experimental curve
outside the KT critical region (this will be seen below).
One calculation going beyond the approximate
KT resistance expression was a simulation of the
sine--Gordon formulation of the Coulomb gas~\cite{WM8588}.
An indirect connection was found to experiments,
but a direct calculation of the resistance curve is lacking.

In this paper we report results of
Monte Carlo (MC) simulation and finite size scaling analysis
of the linear resistance, from equilibrium vortex dynamics
of the 2D Coulomb gas model for the KT transition.
We first use finite size scaling analysis to extract critical properties
from our MC data for the linear resistance,
and we find excellent agreement with
scaling properties of the KT transition.
Secondly, we show that under certain conditions,
our results for the linear resistance can closely fit
the experimental curve above the transition temperature.

Let us first consider scaling properties of the linear resistance
at a superconducting transition driven by vortex fluctuations,
like the KT transition or the vortex glass transition.
At the critical temperature $T_c$ the correlation length diverges:
$\xi \rightarrow \infty$.
Dynamic quantities like the resistance
depend on the correlation time $\tau$, which
diverges as $\tau \sim \xi^z$, where $z$ is the dynamical exponent.
Voltage is related to the rate of phase slip
of the superconducting order parameter,
$V = \hbar/(2e) {\dot \phi} \sim \xi^{-z}$.
The linear resistance is related to the
equilibrium voltage fluctuations according to the
Nyquist formula~\cite{Reif,APY,2DVG}
\begin{equation}
R = \frac{1}{2 T} \int_{-\infty}^{+\infty} dt\ \langle V(t) V(0) \rangle,
\label{Nyquist}
\end{equation}
and thus the resistance scales as $R \sim \xi^{-z}$.

Now we specialize to two dimensions.
The superconducting transition is now a KT transition,
where confined neutral vortex pairs start to deconfine,
so that ``free'' vortices appear.  Here we will assume the value
$z=2$~\cite{AHNS,Teitel}, which describes free vortex diffusion.
Hence the resistance is expected to scale like
\begin{equation}
R \sim \xi^{-2}.
\label{Rscaling}
\end{equation}
This scaling form, here obtained from general scaling
arguments, is the same as from the usual KT theory (see below).
Simulations are done in finite systems,
and the diverging correlation length at $T_c$ is cut off
at the system size $L$.
Precisely at $T_c$ the resistance is therefore
expected to obey $R \sim L^{-2}$, or $L^2 R = A$,
where $A$ is a (nonuniversal) constant.

As an aside comment, we note that
another way of introducing a finite length scale at the transition
is to apply a finite current density $J$.
The finite current length $\xi_J$, defined by
$\xi_J J = k_B T$~\cite{VG},
is the length beyond which the diverging correlations at the
transition are driven out of equilibrium by the Lorentz force.
The resulting nonlinear response obeys
$E = R(J)J \sim \xi_J^{-z}J \sim J^{z+1} = J^3$,
which is the usual KT result, here obtained within a general scaling
approach.

We can extract the leading finite size correction to scaling
of the resistance by using known results from KT theory.
The correlation length $\xi$ diverges as~\cite{PM}
\begin{equation}
\xi^{-2} \sim
\left[ g \frac{4 \pi z}{\epsilon T} \right]^
{1/\left(1 - \frac{1}{4 \epsilon T}\right)} ,
\end{equation}
for $T \rightarrow T_c^+$, where $g$ is a constant, $1/\epsilon$ is a
dielectric function proportional to the renormalized superfluid density,
and $z$ is the fugacity.
The Bardeen--Stephen relation~\cite{BS},
$R/R_N = 2 \pi \zeta^2 n_F$,
ties the resistance to the density of free vortices $n_F$,
where $R_N$ is the normal state resistance and
$\zeta$ the Ginsburg--Landau length,
and $n_F$ is in turn related to $\xi$ according to the
Debye--H\"uckel relation~\cite{PM}, $\xi^{-2} = 2\pi n_F/\epsilon T$.
Eliminating $n_F$ gives back the scaling relation,
$R \sim \xi^{-2}$, in agreement with Eq.~(\ref{Rscaling}).
The leading correction to scaling is known for
$1/\epsilon$ \cite{WM88}:
$1/\epsilon(L) = 1/\epsilon(\infty) (1 + 1/(2\ln(L)+C))$.
Combining these relations gives
again a finite size scaling relation for $R$,
similar to $L^2 R = A$,
but now containing the leading logarithmic correction:
\begin{equation}
L^2 R \left(1 + \frac{1}{4 \ln(L) + C} \right) = A,
\label{LogCorrection}
\end{equation}
where $C$ is an unknown constant.
We also need the standard KT form of the resistance~\cite{PM}.
Using the temperature dependence of $\epsilon$ from KT theory,
$\xi$ becomes
$\xi \sim \exp({\rm const}/\sqrt{T-T_c})$,
which gives
\begin{equation}
R \sim e^{-{\rm const}/\sqrt{T-T_c}},
\label{KTresistance}
\end{equation}
expected to be valid in the KT critical scaling region.

In the simulation we use the
2D Coulomb gas model. We consider a square lattice with
$N=L^2$ sites with periodic boundary conditions.
The grand partition function is
$Z = Tr_{\{n_j\}} \exp (-H/T)$,
with
\begin{equation}
H = \frac{1}{2} \sum_{i,j} n_i n_j G_{ij} - \mu \sum_j n_j^2,
\label{H}
\end{equation}
where the trace is over $n_i=0, \pm 1$, where $n_i=0$ means no vortex
and $n_i = \pm 1$ means one vortex of vorticity $\pm$ at site $i$.
The parameter $\mu = -E_c$ is a ``vortex chemical potential''
and $E_c$ is the vortex core energy.
We only allow configurations of zero total vorticity.
The vortex--vortex interaction is a lattice Green's function,
$G({\bf r})=(2\pi/N)\sum_{{\bf k} \ne0} \exp(i{\bf k} \cdot {\bf r}) /
(4 - 2\cos k_x - 2\cos k_y)$,
which varies like $-\ln r$ on large distances
(the lattice spacing is set to $a=1$).

Now we will describe the simulation.
Following Ref.~\onlinecite{Teitel},
our MC algorithm consists of attempts to insert
near neighbor $(n=+1, n=-1)$ pairs, on randomly chosen
lattice positions, and with random orientation,
and we test for acceptance according to the Metropolis algorithm.
We typically discard $10^4$ initial sweeps through the system
to approach equilibrium,
and then take measurements during up to $10^6$ sweeps close to $T_c$,
and fewer away from $T_c$.
We compute the linear resistance in the simulation by evaluating the
Nyquist formula, Eq.~(\ref{Nyquist}), as a sum over
discrete time steps~\cite{APY,2DVG}.
The voltage $V(t)$ is given by $\pm 1$
(in suitable units), if at MC time $t$
insertion of a vortex pair with orientation $+- (-+)$ is accepted,
and $V(t)=0$ otherwise.
To equate MC time to real time assumes heavily overdamped dynamics,
which should be fulfilled near the transition.

Now we turn to the results.
In Figure~\ref{fig1} we determine the transition temperature from
finite size scaling of MC data for the linear resistance.
The vortex chemical potential was here chosen to
$\mu=0$, which gives the Coulomb gas corresponding
to the 2D XY model in the Villain approximation.
According to Eq.~(\ref{Rscaling})
data for $L^2R$ for different lattice sizes $L$ should be
independent of system size at $T=T_c$.
This means that $T_c$ is where the curves in the figure cross.
Note that this scaling procedure assumes initial knowledge
of the dynamical exponent, $z=2$.
Fig.~\ref{fig1}~(a) is based on Eq.~(\ref{Rscaling}) and does not
contain corrections to scaling.
Fig.~\ref{fig1}~(b) is based on Eq.~(\ref{LogCorrection})
which contains the lowest logarithmic correction to scaling.
The constant $C$ in Eq.~(\ref{LogCorrection}) was adjusted to
as closely as possible have all curves cross at a single point.
Comparing (a) and (b) shows that including the correction
significantly improves the fit, and gives the estimate
$T_c \approx 0.218$ for $\mu=0$.
Error bars on the data points have been left out because
they are much smaller than the symbol size.

Figures~\ref{fig1}~(a) and (b) show that the resistance obeys
the expected scaling form $R \sim L^{-2}$ at
the transition, $T=T_c$.
This also means that in the superconducting phase below $T_c$,
the resistance scales to zero with system size.
This is possible in the Nyquist formula~(\ref{Nyquist}), despite that
the equal--time voltage correlation, $\langle V^2(0) \rangle$,
gives a finite, roughly size independent, contribution to the
resistance at all finite temperatures, also below $T_c$.
What happens is that below $T_c$ the
unequal--time correlations on average cancel the equal--time correlations.
This cancellation is incomplete above $T_c$,
and in finite systems also below.
Within the KT vortex unbinding picture,
below $T_c$ all vortices are bound in neutral pairs.
Creation of a pair gives a burst of voltage that
contributes to the equal--time
voltage correlation, but if the pair is later annihilated a canceling
unequal--time correlation results.
Above $T_c$ some pairs unbind, the cancellation becomes incomplete,
and the resistance is finite.

We wish to compare the present approach of using the linear
resistance to locate the critical point, to the often used
method of locating the universal jump in the superfluid density $\rho_s$.
In the Coulomb gas $\rho_s$ corresponds to
the dielectric response function $\rho_s \sim 1/\epsilon(k \rightarrow 0)$,
where $1/\epsilon({\bf k}) =
1 - 2 \pi/(k^2 T N) \langle |n({\bf k})|^2 \rangle$,
where $n({\bf k})$ the the Fourier transform of the vortex density.
The strong size dependence of $R \sim L^2$ produces much stronger
splay of data away from $T_c$, than the universal jump
which has size dependence $\rho_s \sim L^0$.
This makes locating $T_c$ easier.
Furthermore, use of $1/\epsilon(\bf k)$ is in practice limited to the
smallest available nonzero wave vector, $k=2\pi/L$, while $R$ is for $k=0$.

Now we will compare resistance data to the experimental curve
over a finite temperature interval above $T_c$.
In order to do this we first employ a crucial empirical correction
to our MC results.
Our discrete MC dynamics fails in the limit of high temperature,
because the voltage fluctuations saturate at high $T$
when most MC moves are accepted,
which incorrectly makes the resistance vanish like
$R \sim 1/T \rightarrow 0$ according to Eq.~(\ref{Nyquist}).
Instead of approaching a constant at high $T$, we assume an
additional factor
of $T$ in the vortex velocity squared, as for the temperature dependence
of a diffusion process according to the Einstein relation~\cite{Reif}.
This correction cancels the factor $1/T$ in the Nyquist formula,
and makes $R$ approach a constant at high $T$.
The correction is negligible in the critical region, compared to
the strong exponential temperature dependence in Eq.~(\ref{KTresistance}).

Figure~\ref{fig2}~(a) shows MC data for the resistance curve,
together with the experimental curve.
Note that the temperature scale is here the Coulomb gas temperature,
$T=T^{\rm CG}$, which differs from the physical temperature by the
temperature dependence of the unrenormalized superfluid density
$\rho_{s0}$:
$1/T^{\rm CG} = 2 \pi \rho_{s0}/ k_B T^{\rm physical}$~\cite{PM}.
The only parameters involved in the plot are
$T_c$, which is determined separately
by the above scaling approach, and the normalization
constant $R_N$, which is adjusted to obtain agreement with the
experimental curve at high $T$.
MC data for $\mu=0$ is seen to deviate strongly from
the experimental curve.
To get better agreement we tried
different values of the vortex chemical potential $\mu$.
We obtain best agreement for the value $\mu=-0.8$.
The critical temperature for $\mu=-0.8$ is very close to the
upper limit $T_c=1/4$, and the value $T_c=1/4$ is used in the plot.
Further decreasing $\mu$ gives a worse fit.
Figure~\ref{fig2}~(b) is a KT plot with $1/(T/T_c-1)^{-1/2}$ on the x-axis.
The KT form given by Eq.~(\ref{KTresistance}) corresponds to
a straight line in this plot.

Let us discuss this result.
By adjusting the vortex chemical potential to
$\mu=-0.8$, our data for the resistance is in close
agreement with the experimental curve
over many decades, and over a broad temperature range.
The agreement is not perfect, as indicated by the finite size
effects near $T_c$ in Fig.~\ref{fig2}, but the
agreement is clearly better than a fit to KT theory
outside the KT critical region.
The value $\mu=-0.8$ seems perhaps unexpectedly small,
compared to e.g.\ the 2D XY model which has $\mu \approx 0$.
The corresponding Coulomb gas
model has very small vortex density, and it is very close to the
KT line of fixed points.
One possible interpretation is that
this is actually an effective vortex model with renormalized parameters.
For example, vortex crystallization in superconducting films was
suggested recently by Gabay and Kapitulnik~\cite{Kapitulnik}.
Our Coulomb gas model could describe logarithmically
interacting vacancies with big core energy.

We have also studied the effect of adding
random vortex pinning, modeled as a random site energy for the vortices.
This shifts $T_c$ downwards, but
gives exactly the same scaling properties at the new $T_c$.
This directly verifies the
expectation that weak disorder is
indeed irrelevant at the KT transition,
in agreement with the Harris criterion, $\nu > 2/d$~\cite{Harris,MW}.

In summary, our results from finite size scaling of MC data for the linear
resistance of the 2D Coulomb gas are in excellent agreement
with scaling properties of the KT transition.
This shows that Monte Carlo dynamics is in the right
dynamical universality class,
and allows accurate determination of the critical temperature.
Furthermore, by tuning the vortex chemical potential,
our results can closely reproduce the
experimental universal resistance curve over many decades,
and the agreement outside the KT critical region is
much better than a simple fit to the KT theory.

We acknowledge stimulating discussions with
S.~M.\ Girvin, P.\ Minnhagen, H.~J.\ Jensen, and S.\ Teitel.
This work was supported by grants from the Swedish Natural
Science Research Council (NFR).

\begin{figure}
\caption{Location of the critical temperature by finite size scaling
of Monte Carlo data for the linear resistance $R$.
The critical temperature is
located where data for different system sizes $L$ intersect.
In (a) is shown data without the logarithmic correction
to scaling,
and in (b) is the same data including the logarithmic correction.
}
\label{fig1}
\end{figure}

\begin{figure}
\caption{
Resistance curves from
Monte Carlo simulations for different values of
the vortex chemical potential $\mu$,
together with the experimental curve.
The resistance is plotted vs.\ $T$ in (a),
and vs.\ $(T/T_c-1)^{-1/2}$ in (b) as suggested by
Eq.~(\protect\ref{KTresistance}).
}
\label{fig2}
\end{figure}

\end{document}